\documentclass[modern, a4paper, times,preprint2]{aastex62}
\usepackage{natbib}

\usepackage{amsmath}
\usepackage{graphicx}
\usepackage{natbib}
\received{,}
\revised{,}
\accepted{}
\submitjournal{ApJ}

\shorttitle{Heliosphere in the ENA spectra}
\shortauthors{Czechowski et al.}

\begin{document}
\title{Heliospheric structure as revealed by the 3 -- 88 keV H ENA spectra}

\correspondingauthor{M. Bzowski}
\email{bzowski@cbk.waw.pl}

\author[0000-0002-4441-5377]{A. Czechowski}
\affil{Space Research Centre PAS (CBK PAN)\\ 
Bartycka 18 A, 00-716 Warsaw, Poland}

\author[0000-0003-3957-2359]{M. Bzowski}
\affil{Space Research Centre PAS (CBK PAN)\\ 
Bartycka 18 A, 00-716 Warsaw, Poland}

\author[0000-0002-4173-3601]{J.M. Sok{\'o}{\l}}
\affil{Space Research Centre PAS (CBK PAN)\\ 
Bartycka 18 A, 00-716 Warsaw, Poland}

\author[0000-0002-5204-9645]{M.A. Kubiak}
\affil{Space Research Centre PAS (CBK PAN)\\ 
Bartycka 18 A, 00-716 Warsaw, Poland}

\author[0000-0001-7867-3633]{J. Heerikhuisen}
\affil{Department of Space Science\\
and Center for Space Plasma and Aeronomic Research\\
University of Alabama in Huntsville\\
Huntsville, AL, USA}

\author[0000-0001-7240-0618]{E.J. Zirnstein}
\affil{Department of Astrophysical Sciences\\
Princeton University\\
Princeton, NJ, USA}

\author[0000-0002-6409-2392]{N.V. Pogorelov}
\affil{Department of Space Science\\
and Center for Space Plasma and Aeronomic Research\\
University of Alabama in Huntsville\\
Huntsville, AL, USA}

\author[0000-0002-3737-9283]{N.A. Schwadron}
\affil{Space Science Center and Department of Physics\\
University of New Hampshire\\
Durham, NH, USA}

\author[0000-0003-1703-7777]{M. Hilchenbach}
\affil{Max-Planck-Institut f{\"u}r Sonnensystemforschung\\
G{\"o}ttigen, Germany}

\author[0000-0003-3951-0043]{J. Grygorczuk}
\affil{Space Research Centre PAS (CBK PAN)\\ 
Bartycka 18 A, 00-716 Warsaw, Poland}

\author[0000-0002-4642-6192]{G.P. Zank}
\affil{Department of Space Science\\
and Center for Space Plasma and Aeronomic Research\\
University of Alabama in Huntsville\\
Huntsville, AL, USA}

\begin{abstract}
Energetic neutral atoms (ENA) are an important tool for investigating the structure of the heliosphere. Recently, it was observed that fluxes of ENAs (with energy $\le$ 55 keV) coming from the upwind and downwind regions of the heliosphere are similar in strength. This led the authors of these observations to hypothesize that the heliosphere is bubble-like rather than comet-like, meaning that it has no extended tail. We investigate the directional distribution of the ENA flux for a wide energy range (3--88 keV) including the observations from IBEX  (Interstellar Boundary Explorer), INCA (Ion and Neutral Camera, on board Cassini), and HSTOF (High energy Suprathermal Time Of Flight sensor, on board SOHO, Solar and Heliospheric Observatory). An essential element is the model of pickup ion acceleration at the termination shock (TS) proposed by Zank. We use state of the art models of the global heliosphere, interstellar neutral gas density, and pickup ion distributions. The results, based on the "comet-like" model of the heliosphere, are close in flux magnitude to ENA observations by IBEX, HSTOF and partly by INCA (except for the 5.2-13.5 keV energy channel). We find that the ENA flux from the tail dominates at high energy (in agreement with HSTOF, but not INCA). At low energy, our comet-like model  produces the similar strengths of the ENA fluxes from the upwind and downwind directions, which, therefore, removes this as a compelling argument for a bubble-like heliosphere.
 
\end{abstract}



\section{Introduction} \label{sec:intro}
\noindent
Energetic Neutral Atoms (ENAs) considered in this work are created due to multi-stage 
processing of interstellar neutral (ISN) atoms inside the heliosphere. Unlike interstellar 
ions, ISN atoms freely penetrate deep inside the termination shock (TS), where some of them are 
ionized by photoionization and charge exchange with protons from the solar wind (SW). The 
ionized ISN atoms form a suprathermal sub-population of the SW, known as pickup ions (PUIs) 
\citep{holzer_axford:71a,vasyliunas_siscoe:76}. Due to a sequence of interactions with the TS, 
some of the PUIs are accelerated to energies of dozens of keV and larger before being convected 
into the inner heliosheath, i.e., the region of SW plasma between the TS and heliopause (HP) 
\citep{lee_etal:96a,zank_etal:96b,zank_etal:10a,kumar_etal:18a}. In the inner heliosheath, the 
PUIs maintain their suprathermal energy and flow with the shocked SW plasma until they 
are converted into ENAs by neutralization, predominantly through resonant charge exchange with 
ISN H atoms that have penetrated inside the heliopause. These ENAs propagate freely in all 
directions, carrying information about the ion distributions in their inner heliosheath source 
region.

The observation of ENAs is presently one of the most important means for diagnosing the global 
heliosphere \citep{gruntman:97}. The motion of the Sun relative to the interstellar medium affects the 
distribution of the primordial seed population of ENAs, i.e., the ISN atoms inside the TS, and 
consequently the production of PUIs. The acceleration of PUIs to ENA energies and their transport 
downstream of the shock are determined by the heliospheric plasma flow geometry. Therefore, to 
understand the relationship between the ENA flux distribution and the structure of the heliosphere we 
must harness theoretical models of (1) the ISN H distribution and the production of PUIs inside TS, (2) 
the plasma flow in the inner heliosheath, where the heliospheric ENAs observed at Earth's orbit are 
produced, and (3) the PUI distribution at the TS and its evolution as it is convected through the 
heliosheath.

At present, theoretical models of the heliosphere give only an approximate picture. In particular,
the thickness of the inner heliosheath (an important parameter for the ENA distribution) is
overestimated by all MHD and MHD-kinetic numerical solutions. Moreover, Voyager observations at the TS imply that
the highly nonthermal ion component have a large effect on plasma dynamics in this region, not
yet fully taken into account in most state-of-art models. We cannot therefore aim at a precise 
modeling of the ENA distribution, while following theoretical models uncritically.  

Altogether, we address the following topics:

(A) The energy spectrum and directional distribution of the hydrogen ENAs over a wide energy range 
including the high energy (above $\sim$ 40 keV) ENA. This permits us to link with the Voyager LECP 
energetic ion measurements (E$\ge$28 keV for V1 and E$\ge$40 keV for V2) and test the theoretical model of 
the flux intensity and the energy spectrum of the pick-up ions at the TS proposed by 
\citet{zank_etal:96b, zank_etal:10a}. The energy range extends from
3~keV (the highest bin of IBEX High) to 88~keV (maximum energy of HSTOF) and includes the
observations by INCA.

(B) Tail/nose asymmetry of the ENA flux in the heliosphere, in particular the energy dependence
of the tail/nose ENA flux ratio. Our aim here is to find out to what extent this ratio can be 
regarded as a signature of the heliosphere with the extended tail (the "comet-like" heliosphere)
\citep{parker:61a,baranov_etal:91}.
In the present work we do not employ a fully time-dependent model of the heliosphere, so
we cannot discuss the important issue of time dependence of the ENA flux coming from the tail
direction \citep{dialynas_etal:17a}. However, this topic was recently addressed by 
\citet{schwadron_bzowski:18a}. 

We restrict our simulations to the vicinity of the ecliptic plane, to stay within the observation
region of HSTOF. 

Our simulations are organized as follows. The locations of the TS and the HP
and the plasma flow in the inner heliosheath are taken from the time-stationary Huntsville model
of the heliosphere \citep{heerikhuisen_pogorelov:10a}, run with the currently best 
parameters of interstellar gas populations \citep{mccomas_etal:15b, bzowski_etal:15a, 
kubiak_etal:16a}, and interstellar magnetic field \citep{zirnstein_etal:16b, frisch_etal:15c}.
The Huntsville model does not explicitly include a separate (PUI) component: the plasma fluid
in the simulation should be considered as the bulk SW and PUI mixture.  

We start from the hot model of ISN H distribution inside the heliosphere 
\citep{tarnopolski_bzowski:09} using an observation-based model of the time- and latitude 
evolution of radiation pressure \citep{kowalska-leszczynska_etal:18a}, photoionization rate 
\citep{bzowski_etal:13a}, and of the SW speed and density \citep{sokol_etal:13a}, needed to 
calculate the charge-exchange ionization rate. From the simulated ISN H distribution between 
the Sun and the TS, we calculate the density of PUIs arriving at the TS in the ecliptic plane 
\citep{sokol_etal:19b}, for two selected phases of the solar cycle (Fig.~\ref{fhpui}). The choice 
of parameters for this model is presented in Appendix A.

As a next step, we use the Zank acceleration theory of the transmission and reflection of PUI components at the TS to obtain the PUI spectra just beyond the TS \citep{zank_etal:96b,zank_etal:10a}. The example of the proton spectrum corresponding to the TS parameters encountered by Voyager~2 is shown in Figure \ref{fv2}, with contributions of the bulk SW protons and of the transmitted and reflected PUIs. We then follow the convection and gradual decharging of these ions as they propagate within the IHS plasma. We use the heliosheath distribution of ISN H, the TS location, and the plasma flow from the Huntsville heliosphere model. 

The termination shock transition observed by Voyager~2 consisted of a narrow subshock and an extended (0.7~au) precursor \citep{florinski_etal:09a}. In the Huntsville model, where the plasma is treated as a single fluid, this transition appears as a single shock, with the strength combining the subshock and the precursor strengths. To apply the Zank et al. theory, we have to estimate the strength of the subshock. We make a simple assumption that the subshock strength is lower than the shock strength in the Huntsville model by a constant fraction. Its value (0.68) is chosen by the requirement that the resulting subshock strength at the point of Voyager~2 crossing the simulated shock is equal to the TS strength determined from Voyager~2 observations

Finally, we simulate the ENA flux along the ecliptic plane in the energy bands corresponding to IBEX, INCA, and HSTOF measurements. The comparison of the simulated ENA flux with the observations by IBEX \citep[3--6~keV, midpoint 4.3~keV,][]{schwadron_etal:14b}, by INCA \citep[at 5.2--13.5~keV and 35--55~keV, ][]{dialynas_etal:17a}, and by HSTOF \citep[58--88 keV,][]{hilchenbach_etal:98a, czechowski_etal:06b} is presented in Figure \ref{ftn} (tail and nose directions) and Figure \ref{fcmp} (all directions near the ecliptic). Figure~\ref{ftn} includes an additional data point for IBEX (2--3.8~keV, midpoint 2.7 keV). 

The detailed description of the models used in our simulations is given in Sections 2 to 5. Specifically, Section 2 presents the global model of the heliosphere, Section 3 our method of obtaining the pick-up protons density distribution upstream of the TS, Section 4 the theory of acceleration of PUIs at the TS and the resulting energetic ion spectra, and Section 5 the conversion of the energetic protons to ENAs. Section 6 provides information about the ENA flux data used in this work. Our results and conclusions are summarized in Section 7. In Appendix A we include more details about the parameters of the interstellar medium and the heliosphere. Appendix B describes the toy model of the ENA tail/nose flux ratio. In Appendix C we present some MHD+neutral models of the heliosphere with two-funnel topology, corresponding to the case of strong interstellar magnetic field.

\begin{figure}
\centering
\includegraphics[width=0.6\textwidth]{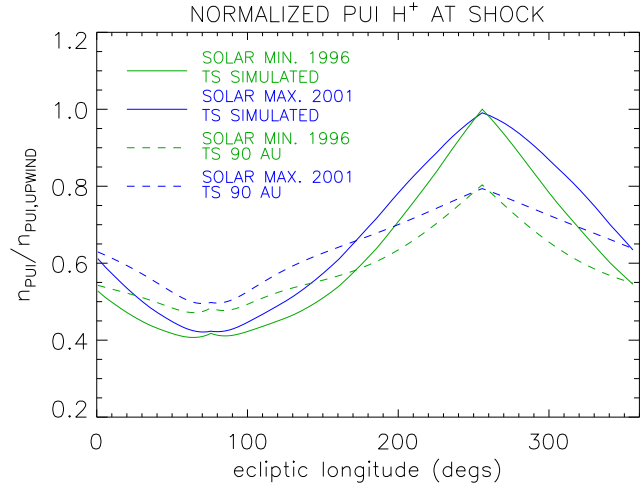}
\caption{Variation of the normalized H$^+$ PUI density along the TS in the ecliptic plane following 
from the nWTPM model. The solid lines correspond to the TS location derived from the Huntsville 
model, and the dashed lines to the hypothetical case of the Sun-centered spherical TS with a radius 
of 90~au. For all the curves of the same kind, the density is normalized to the same upwind value 
(longitude $255\degr$) during the 1996 solar minimum. The PUI normalized density distribution is 
weakly sensitive to the phase of the solar cycle (green vs blue), but much more sensitive to the 
geometry of the TS (solid vs dashed). The absolute values of the PUI density along the TS are 
different for the spherical Sun-centered and the simulated shock locations.}
\label{fhpui}
\end{figure}

\begin{figure}
\centering
\includegraphics[width=0.5\textwidth]{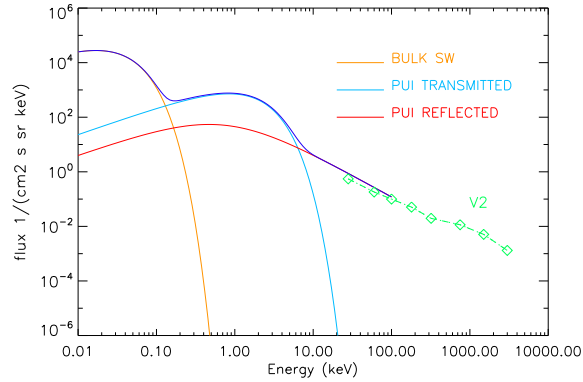}
\caption{Simulated energetic proton spectrum downstream from the TS obtained from the model of 
\citet{zank_etal:10a} for the shock parameters as observed by Voyager 2, compared with Voyager 2 / 
LECP $Z\ge 1$ ions measurements \citep{giacalone_decker:10a}. The simulated spectrum is a 
superposition of two Maxwell-Boltzmann functions for the bulk SW and transmitted PUI populations, and 
a kappa function with $\kappa=1.6$ for the reflected PUI population. The dark blue line is the sum of all components.}
\label{fv2} 
\end{figure}

\begin{figure}
\centering
\includegraphics[width=0.5\textwidth]{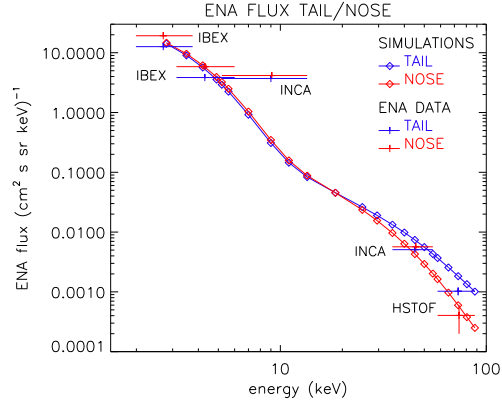}
\caption{Simulated H ENA spectrum from the tail and the nose directions compared with observations by IBEX (2.7 and 4.3 keV), INCA (5.2--13.5 keV and 35--55 keV) and HSTOF (58--88 keV). An antisunward-looking observer is located at 1~au. The ENA data are the same as in Fig. \ref{fcmp}, averaged in ecliptic longitude over the nose $255\degr\pm25\degr$ and the tail $75\degr\pm25\degr$ regions, respectively. The horizontal bars correspond to the energy ranges of the observations, the vertical bars are the measurement uncertainties estimates. For the IBEX data points the vertical bars do not represent the errors, which are too small to visualize. The simulated spectra from the tail and the nose regions agree with each other for the energies up to $\sim 20$ keV (simulations) or $\sim 40$ keV (observations), and diverge for higher energies. The divergence can be explained by the rapid fall-off in the charge-exchange cross section at high energy, which effectively extends the production region of high energy ENAs in the heliotail. Note that the only measurement that markedly differs from the model ENA spectrum, and also from the observed IBEX Hi spectrum, is the INCA 5.2--13.5 keV range. }
\label{ftn}
\end{figure}

\begin{figure}
\centering
\includegraphics[width=0.5\textwidth]{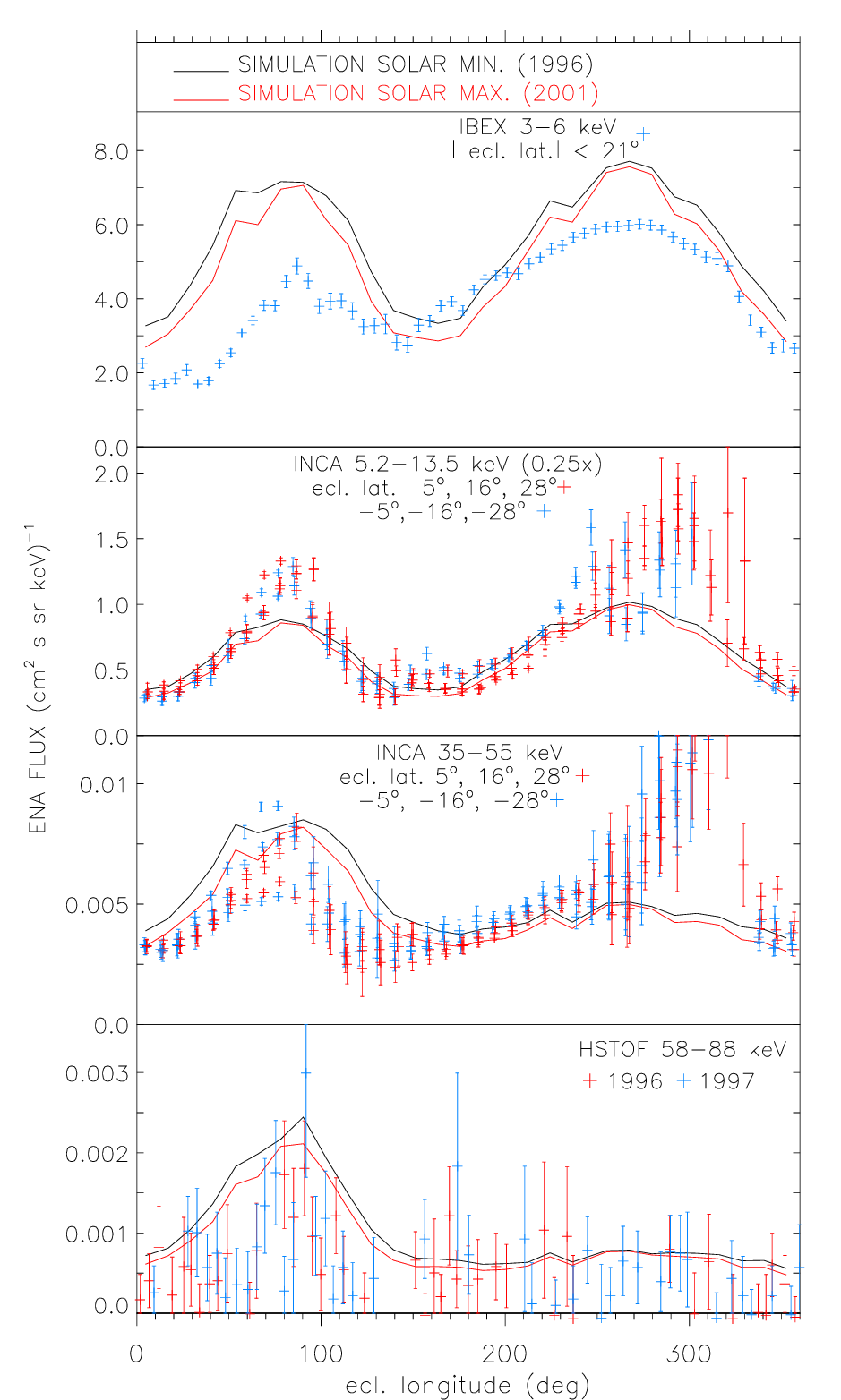}
\caption{Simulated H ENA flux near the ecliptic plane compared with observations. In the second panel 
(the INCA 5.2--13.5 keV data), the observed flux was scaled down by a factor 0.25 to facilitate 
comparison with the simulations. The H ENA flux was observed by IBEX \citep{schwadron_etal:14a} 
between 2009--2014, INCA in two energy bands (digitized from Figure 3a in \citet{dialynas_etal:17a}), 
and HSTOF \citep{czechowski_etal:06b}. The simulations, averaged over the energy ranges of the 
measurements, were done using the (time-stationary) Huntsville model of the heliosphere together with 
the nWTPM model results for the PUI distribution during the solar minimum (black) and maximum (red) 
conditions. The INCA data, gathered between 2003--2014, correspond to a range of latitudes, both 
above (red) and below (blue) the ecliptic plane. The HSTOF data come from the first two years of 
operation, with favorable observing conditions: 1996 (red) and 1997 (blue).}
\label{fcmp}
\end{figure}

\section{Model of the heliosphere}
\noindent
The locations of the TS and the HP and the plasma flow in between are obtained from the time-stationary Huntsville model. The model combines an MHD description of the interaction of solar and interstellar plasmas with a kinetic description of neutral hydrogen atoms \citep{pogorelov_etal:08b,heerikhuisen_pogorelov:10a}. The SW and interstellar plasmas are described as a single fluid under ideal MHD equations. The MHD equations are coupled to neutral hydrogen through mass, momentum, and energy source terms via photo-ionization and charge-exchange. The Boltzmann equation is solved using a Monte Carlo approach in order to solve for the neutral hydrogen distribution in phase space. We use the MHD solution for (1) estimating the TS distances, (2) retrieving the neutral H density in the heliosheath, and (3) retrieving the plasma streamlines in the heliosheath. The other information needed for our simulations is derived from the nWTPM model for the neutral H, the model of the PUI distribution upstream from TS \citep{sokol_etal:19b}, and the Zank's theory for the PUI temperatures across the shock.

The boundary conditions for the model are specified as follows. At 1~AU, the SW plasma density is 5.74 cm$^{-3}$, temperature is 51,100~K, flow speed is 450~km~s$^{-1}$, and the magnetic field radial component is 37.5 $\mu$G, all independent of heliolatitude. These values are then advected to the simulation's inner boundary at 10~AU assuming adiabatic expansion. Neutral hydrogen atoms generated in the heliosphere outside of the TS are adopted in our calculation directly from the Huntsville model, assuming for the plasma a kappa distribution in the inner heliosheath and a Maxwell-Boltzmann distribution outside the heliopause.

In the unperturbed Very Local Interstellar Matter (VLISM), the strength of magnetic field was adopted 
as 2.93 $\mu$G, pointing towards ecliptic (longitude, latitude) = $(227.28\degr, 34.62\degr)$ 
\citep{zirnstein_etal:16b}. The VLISM temperature and speed were adopted 7500~K and 25.4~km~s$^{-1}$, 
respectively \citep{mccomas_etal:15b}, the proton density 0.09 cm$^{-3}$, and the H density 
0.154~cm$^{-3}$ \citep{zirnstein_etal:16b}.

Along the Voyager 1 (2) directions, the model gives the distance to the heliopause $r_{\text{HP}} = 118 (115)$~AU, to the termination shock $R_{\text{TS}} = 74 (74)$~AU, implying the thickness of the inner heliosheath $L_{\text{IHS}} = 44 (41)$~AU. The observed values are 121 (119)~AU, 94 (84)~AU and 27 (35)~AU, respectively.  

During the past 25 years, the SW flux featured a secular change (in addition to 
quasi-periodic solar cycle variations), with a gradual decrease between $\sim 1990$ and 2010, 
an increased plateau during 2010--2014, and a sharp increase in 2014 (see, e.g., Figure 4 in 
\citet{czechowski_etal:18a} and Figure 1 in \citet{mccomas_etal:18a}). This very likely 
resulted in significant changes to the structure of the heliosphere, in particular in the TS 
distance (depending on the location along the TS). To correctly estimate the thickness of the 
heliosheath from the Voyager 1 (2) measurements of the distance to TS in 2005 (2008) and to HP 
in 2012 (2018) requires taking this time-dependence into account.

\section{Pick-up protons density upstream of TS}
\noindent
The parent ions of the ENAs considered in our simulation are the pick-up protons accelerated at the TS. To determine their distribution, we must know the density of the pick-up protons arriving at TS with the SW.    

Calculation of the PUI densities along the TS was done using the formula \citep{rucinski_fahr:91, rucinski_etal:03, sokol_etal:19b}:
\begin{equation}
F_{\text{PUI}}(\vec{r}_{\text{TS}}, t) = \frac{1}{r_{\text{TS}}^2(\vec{\omega})}\int\limits_{r_0(\vec{\omega})}^{r_{\text{TS}}(\vec{\omega})}n_{\text{H}}(\vec{r}', t) 
\beta(\vec{r}', t)r'^2 d r'
\end{equation}
where $n_{\text{H}}$ is the density of ISN H for the time $t$ at radius-vector $\vec{r}${} from the Sun, and $\beta$ is the ionization rate of ISN H at radius-vector $\vec{r}'$ for a time $t$. The heliocentric radius vector $\vec{r}(r, \vec{\omega})$ is parametrized by its length $r$ and direction (a directional unit vector $\vec{\omega}(\lambda, \phi)$, where $(\lambda, \phi)$ are ecliptic longitude and latitude). Note that $\beta$ is for a solar distance $r$, not for 1~AU, and that it varies with heliolatitude. The integration goes radially from a distance $r_0$ from the Sun to the TS distance $r_{\text{TS}}$ along the direction $\vec{\omega}$. Assuming that PUIs propagate radially and the solar wind speed is independent of solar distance, the density of PUIs at TS is calculated as $n_{\text{PUI}} = F_{\text{PUI}}/v_{\text{SW}}$. This calculation is an approximation where the slowdown of the solar wind due to momentum loading by newly-injected PUIs and the effects of finite propagation time of solar wind from the Sun to the TS are neglected. A discussion of the validity of this approximation is provided by \citet{bzowski_etal:13a}, pages 82--86. 

The densities $n_H$ were calculated adopting the paradigm of the classical hot model of the ISN H 
distribution in the heliosphere \citep{thomas:78}, with modifications to account for the dependence 
of the solar radiation pressure on time and radial velocity due to the evolution of the solar 
Lyman-$\alpha$ emission profile with time \citep{tarnopolski_bzowski:09} and for the variation of 
the ionization rate with time and heliolatitude \citep{rucinski_bzowski:95b,bzowski:03}.

In the hot-model paradigm, the density of ISN H at a location given by a radius-vector $\vec{r}$ and 
time $t$ is calculated by numerical integration of the local distribution functions of the primary 
and secondary populations $f_{\text{pri}}$, $f_{\text{sec}}$ over the three-dimensional velocity space 
\citep{baranov_etal:98a}:
\begin{equation}
n_{\text{H}}(\vec{r}, t) = n_{\text{H,pri}}(\vec{r}, t)+n_{\text{H,sec}}(\vec{r}, t) = 
\int (f_{\text{pri}}(\vec{v}, \vec{r},t) + f_{\text{sec}}(\vec{v}, \vec{r},t)) d\vec{v}.
\end{equation}

These distribution functions strongly vary with time, distance to the Sun, and ecliptic coordinates. 
The local distribution function $f(\vec{v},\vec{r}, t)$ is ballistically connected by atom 
trajectories with the distribution function of ISN H at the TS, assumed to be Maxwell-Boltzmann with 
the parameters discussed in Appendix A. The radiation pressure varies with time, heliolatitude, and 
radial speed of the atom along its trajectory. The model used here is by 
\citet{kowalska-leszczynska_etal:18a}. This results in different velocity vectors of the H atoms in 
their source region in the interstellar medium for identical velocities at $\vec{r}$ but for 
different times $t$. The ionization losses also vary between trajectories of different H atoms 
intersecting at the time $t$ at $\vec{r}$. The atom velocity vectors in the interstellar medium and 
the ionization losses for individual trajectories are calculated by numerical integration of the 
respective quantities along the numerically-tracked trajectories (see Section 2 in 
\citet{bzowski_etal:13b}).

PUIs are produced due to ionization of ISN H by charge exchange with SW protons, by photoionization by solar EUV photons, and by electron-impact ionization, which all vary with the solar distance, heliolatitude, and time \citep{bzowski_etal:13a,sokol_etal:19a}:
\begin{equation}
\beta_{\text{ion}}(\vec{r}, t) = \beta_{\text{prod}}(\vec{r}, t) = 
\beta_{\text{cx}}(\vec{r}, t) + \beta_{\text{ph}}(\vec{r}, t) + \beta_{\text{el}}(\vec{r}, t).
\end{equation}

The SW speed model \citep{sokol_etal:13a} used to calculate the charge-exchange and electron 
ionization rates was also used to compute the PUI flux and density. In this model, the SW speed 
varies with time and heliolatitude but is constant with the solar distance, and the solar wind 
density and the photoionization rate drop with the square of distance to the Sun and vary with 
heliolatitude and time (cf. Fig.20 in \citet{sokol_etal:13a}). The ionization model is described in 
\cite{sokol_etal:19a}.

\section{Acceleration of PUIs at the TS and the energetic ion spectra.} 
\noindent
To simulate the proton distribution in the heliosheath, we invoke the mechanism proposed by 
\citet{zank_etal:96b} and independently by \citet{lee_etal:96a}. The energized ions 
derive from PUIs arriving at the TS. The ions with a high enough perpendicular velocity component 
overcome the electrostatic potential barrier at the shock and are transmitted downstream. The 
remaining ones are initially reflected from the shock and spend some time drifting along the shock 
surface while gaining energy before ultimately crossing the shock downstream. 

The shock strengths along the TS were scaled by a factor of 0.68 relative to the Huntsville 
simulation results. The scale factor was chosen by the requirement that the (rescaled) shock 
strength at the point where the model TS is crossed by Voyager 2 trajectory is equal to the 
value measured by Voyager 2. The result is similar to the subshock strength, predicted by 
models of the termination shock that incorporate distinct ACR and PUI components 
\citep{florinski_etal:09a,donohue_zank:93a}.

Note that the MHD-kinetic model which we use includes neither the ACR nor the PUI component as explicit separate fluids. Were the ACRs included along with their spatial diffusion coefficient, the overall shock would be mediated quite significantly and an extended smooth foreshock would be present \citep{donohue_zank:93a,florinski_etal:09a, zank:15a}. The important point here is that the actual shock with which the PUIs interact will be much weaker than given in the MHD and MHD-kinetic models, because these models simply identify the total jump, which includes the ACR foreshock contribution as well.

Following \citet{zank_etal:96b,zank_etal:10a}, we estimated the fractions of the transmitted and 
reflected ion populations and their average energies. We assumed a simple "filled shell" distribution for the PUIs upstream from the shock (Equation 1 from \citet{zank_etal:10a}). To estimate the fractions 
of transmitted and reflected ions, we used the formula for the critical velocity for specular 
reflection from Equation (8) from \citet{zank_etal:96b}, and Equations (11a) and 
(11b) from \citet{zank_etal:10a}. To calculate the critical velocity for specular 
reflection, we used the radial component of magnetic field at 1 au equal to 30.0~$\mu$G, similar 
to that used in the Huntsville model, and we calculated the local magnetic field vector 
assuming the Parker spiral. Taking 37.5 $\mu$G does not lead to any significant changes.

Average energies of the transmitted and reflected ions were taken from Equation (8) and (10) from 
\citet{zank_etal:10a}. Since we do not know the $L_{ramp}$ magnitude along the TS, 
we replaced it with the local values of the ion inertial length, which were determined based on 
local values of plasma density taken from the Huntsville simulation. The temperature of the bulk 
SW downstream of TS was set to $2\cdot 10^5$ K, in agreement with Voyager 
observations \citet{richardson:08a}. This approach correctly reproduces the observation from Voyager 
2 during its TS crossing (see Figure \ref{fv2}).

The ion spectrum downstream was modeled as Maxwellian distributions for the bulk SW 
and transmitted ions, and a kappa-function for the reflected ions, with the number densities 
and average energy values predicted by the above mentioned formulas from 
\citet{zank_etal:10a,zank_etal:96b}. At each location along the TS where these formulas were 
used, we took appropriate plasma parameters from the Huntsville model simulation. The shock 
strength values were multiplied by the scale factor.  For the kappa function, we used the 
value $\kappa=1.6$, which is the approximate slope of the TS particle distributions observed 
by Voyagers \citep{decker_etal:05a}. The PUI densities along the TS were calculated 
as described in Section 3 (Fig. \ref{fhpui}).

\section{Generation of ENAs and their transport to 1 AU}
\noindent
We assumed that the parent ions for the ENAs are transported from the TS by convection along the plasma flow lines. To calculate the ENA flux arriving at 1 au from a given direction, we considered a segment of the radially-directed line of sight (LOS) between the TS and the HP or the outer boundary of 988~AU. For a set of points along this LOS, we determined the plasma flow line that links this point with the initial point at the TS. The plasma flow lines and plasma densities were taken from the global MHD-kinetic simulation. For a selected value of the ENA energy in the observer's frame, for each point along the LOS we found the parent ion velocities relative to the plasma. Subsequently, moving backward along the flow line, we determined the parent ion velocities at the initial point at the TS. This determination takes into account adiabatic acceleration of ions along the flow line. The amount of adiabatic acceleration was obtained based on the plasma density distribution along the flow line. We did not assess this acceleration from divergence of the flow to avoid calculating numerical derivatives on a relatively sparse grid and the singularity at the shock.

With the ion velocity at the TS calculated, we computed the values of the energetic ion flux implied by our model spectrum at the TS. Simultaneously, we determined the loss factor for these energetic ions during their convection along the flow lines due to their neutralization via charge exchange with the ambient interstellar neutral hydrogen. The density and velocity of the background interstellar neutral hydrogen are taken from the Huntsville model. The velocity-dependent charge exchange cross section was adopted from \citet{lindsay_stebbings:05a}. In this way, we determined the production rate of the ENAs moving towards the observer at selected points of the LOS. These production rates were subsequently integrated along the LOS to yield the ENA flux for this LOS. The losses to the ENAs on the way to the observer were not considered because they are small for our energy range \citep{bzowski:08a, mccomas_etal:12c} and we neglect any hypothetical production of ENAs inside the TS.

The results of our simulations of the ENA flux are shown in Figures \ref{ftn} and \ref{fcmp}. 

\section{ENA flux data}
 \noindent
The ENA flux data used in this study consist of:
 
(1) The IBEX Hi globally distributed ENA flux, that is the flux obtained after removing the 
ribbon contribution. These data come from the period 2009--2014 and were published in 
\citet{schwadron_etal:14a}. We used the flux data from the directions within $\pm 10\degr$ from 
the ecliptic plane, in the energy ranges 3--6 keV and (Fig. \ref{ftn} only) 2--3.8 keV.

(2) The HSTOF hydrogen ENA data obtained during the two first years of operation (1996--1997).
This period of observations is unique, since it combines good quality observations with the 
coverage of all ecliptic longitudes. The HSTOF energy range is 58-88~keV.

HSTOF ENA observations were possible only during the quiet time periods, with low energetic 
ion flux. The years 1996-1997 included a large number of such periods. In 1998, connection 
with SOHO  was temporarily lost. The subsequent four 
years had few quiet time periods. In the year 2003 SOHO was re-oriented, with the result that 
the nose and tail ecliptic longitude sectors became inaccessible for HSTOF.
 
It should be noted that the first publication of the HSTOF ENH flux data 
\citep{hilchenbach_etal:98a} appeared before the in-flight calibration of the instrument, 
with the result that the flux was underestimated by about one order of magnitude, as outlined 
in \citet{hilchenbach_etal:01a}. The HSTOF data which we use here are derived from the 
re-calibrated data with a more stringent quiet time flux threshold and resampled binning time 
periods for the ENA flux averaging (published in \citet{hilchenbach_etal:01a} and for an extended 
period \citet{czechowski_etal:06b}). In particular, a single high-flux data point near $0\degr$ 
ecliptic longitude shown in \citet{hilchenbach_etal:98a} is absent from the present data due 
to the revised data analysis while the several data points for H ENA fluxes emerging from the 
tail region are still present.
 
(3) The INCA ENA data in two energy bands (5.2--13.5 keV and 35--55 keV) have been digitized from Figure 3a in \citet{dialynas_etal:17a}). The data from INCA were collected between 2003 and 2014, with different regions in the sky observed at different times.

\section{Results and Conclusions}
\noindent
Our main result (see Figures \ref{ftn} and \ref{fcmp}) is that the observed heliospheric ENA flux can 
be approximately reproduced, over a wide energy range, by a model combining the time-stationary 
conventional ("comet-like") model of the heliosphere with the model of the energetic proton spectrum 
based on Zank et al. theory (Fig. \ref{fv2}). Over most of the energy range, our simulations give a 
right order of the ENA flux magnitude (except for INCA 5.2--13.5 keV data), and the directional 
(longitudinal) dependence of the ENA flux is similar to that observed by IBEX and HSTOF. The 5.2-13.5 
keV INCA data have the longitudinal dependence similar to our simulations, but there is a discrepancy 
in flux magnitude by a factor of 4.

We obtained this result without any parameter fitting: it is based solely on the parameter values 
available in the literature. In our opinion, we have obtained a good qualitative agreement between 
the data and observations. Therefore, even though the model we have used is a single-fluid plasma 
(albeit with the multi-component distribution function) we believe these simplifications are of minor 
importance. Models including details such as effects of spatial diffusion of ions in the energy range 
of 5--100 keV are, to our knowledge, unavailable so far.

The bimodal nose-tail structure of the ENA flux (both simulated and observed) changes with increasing 
energy to a structure with one-peak in the tail direction. For the simulated flux this result is 
robust against various assumptions on the details of the PUI spectrum. In particular, the evolution 
of the bimodal structure is not affected by the adjustment of the shock strength parameter. Switching 
from the two-peak to the one-peak structure for the simulated flux occurs at $\sim$20 keV (Fig. 
\ref{ftn}). A similar behavior can be derived from a simple ``toy'' model (Appendix B). However, the 
INCA data have the bimodal structure both in the 5.2-13.5 keV and the 35-55 keV energy ranges (Fig. 
\ref{fcmp}). The single-peak structure is observed only by HSTOF (58-88 keV). Our interpretation is 
that switching to a single-peak structure in the ENA flux occurs at the energy higher than predicted 
by our simulations.

The simulated ENA flux is only weakly dependent on the solar phase (minimum or maximum). However, in 
each case the simulation is based on the same stationary model of the heliosphere, so that the effect 
of the solar cycle on the global structure of the heliosphere is not taken into account.

Since we employed the time-stationary model of the heliosphere, we could not address the important 
question of time-evolution of the energetic ions and the ENA fluxes observed during recent years. We 
have only considered the effect of the solar cycle on pick-up proton distribution upstream of the 
termination shock (Figure \ref{fhpui}).

In our simulations we assume that the parent ions of the ENAs of different energies are convected 
with the same plasma flow, described by a single fluid which is a mixture of bulk SW and the PUIs. 
The question is whether the single fluid description may offer an acceptable approximation. For 
example, if the core SW and different PUI components downstream from the TS would be transported 
towards widely separated regions of the heliosheath, our approximation would be invalid, and our 
conclusions unsupported. We think, however, that this possibility is unlikely.

The main support for our opinion comes from observations by the Voyager 2 team \citep{richardson_decker:14a,richardson_decker:15a,decker_etal:15a}. Most of  thermal pressure in the inner heliosheath is in the energy band 5.2--24 keV/nuc, which dominate the 5.2 to 3500 keV ion distribution \citep{dialynas_etal:19a}. The agreement between the plasma velocity measurements by the plasma instrument PLS on board Voyager 2 and the velocity estimations based on energetic particles anisotropy observed by LECP is consistent with the energetic particles being convected with the plasma flow for most of the observation period (2008--2014). The apparent discrepancy between the two observed in the 2009.3--2010.5 period can be explained as due to a contribution of heavy ions \citep{richardson_decker:14a,richardson_decker:15a}. The streaming of energetic ions occurs only during the period 2012.7--2013.3, which coincides with an abrupt fall in the intensity of the energetic ions \citep{richardson_decker:14a,richardson_decker:15a,decker_etal:15a}. Our assumption that the plasma and the energetic particles flow together is, therefore, consistent with observations, with the exception of the short period during which the energetic particle flux is very low.

The differential flow between energetic ions of different energies might arise as a consequence of ion diffusion and drift in the heliosheath magnetic field. In the Huntsville model and our ENA simulation the ion diffusion and drifts are neglected. This is consistent with the estimations by \citet{florinski_etal:09a,mostafavi_etal:17b} which imply that the mean free paths for the ions in our energy range are much smaller than the size of the heliosheath. The differential flow does not then occur. To our knowledge, the available heliospheric models that include diffusion do not predict a qualitatively different structure of the heliosphere than the standard models \citep{fahr_etal:00, scherer_ferreira:05b, malama_etal:06, guo_etal:19a}.

Note that our ENA calculations do not assume that the spatial ion distributions are independent of energy. This is because we calculate the ion losses to neutralization along the plasma flow lines starting from the TS, separately for each energy. In this way our simulation goes beyond a single-fluid model.

Our simulations, and to some extent the observations, demonstrate that the ratio between the ENA flux from the tail and the nose directions is energy-dependent (Figures \ref{ftn}, \ref{fcmp}).

In the heliosphere with an extended tail this can be understood as a consequence of energy dependence of the ENA production rate. The cross section for charge-exchange between the energetic proton and the neutral hydrogen atom decreases rapidly with the collision energy \citep{lindsay_stebbings:05a}, by two orders of magnitude over the combined energy range of IBEX-Hi, INCA, and HSTOF (0.7 to 88 keV). Because of losses by conversion into ENAs, the protons energized at and convected from the TS do not fill the inner heliosheath uniformly. The effective source region for the high-energy ENAs observed by HSTOF (58--88 keV, proton loss rate low) extends to larger distances from the TS than the source region for the lower-energy ENAs observed by INCA (5--55 keV). Therefore, the fraction of the heliotail ENAs in the INCA data is smaller than in the HSTOF data, which can be the reason for the difference between the observations by HSTOF (maximum ENA flux from the heliotail) and INCA (similar flux from the nose and tail directions). This qualitative argument has been positively verified using simplified comet-like models of the heliosphere \citep{czechowski_etal:18a}, and further confirmed by the present simulation. The INCA 35--55 keV data and the HSTOF data show that the switch to the one-peak structure of the ENA flux occurs above $\sim 50$ keV, while the Huntsville model-based calculation implies a lower threshold of $\sim 20$~keV (Figures \ref{ftn} and \ref{fcmp}). 

The ENA flux peak dominating at high energy we interpret as a signature of the heliotail.
The peak appears not at the $\sim 75\degr$ "anti-nose" position but is shifted to about 90\degr. 
This shift is visible both in the simulation and in the ENA data from HSTOF. In the simulation 
it reflects the interstellar magnetic field direction, which is inclined at $\sim$40$\degr$
relatively to the direction of the solar motion. 

An alternative to the "comet-like" heliosphere may emerge in the limit of a strong interstellar 
magnetic field, when the magnetic pressure dominates the ram and thermal pressures. According to 
one of the models proposed by \citet{parker:61a}, the heliosheath outflow forms two tubes parallel 
and antiparallel to the interstellar magnetic field and the downwind tail is missing. However, 
this type of the heliospheric structure is only confirmed by numerical simulations where the 
interstellar magnetic field direction happens to be parallel to the solar motion 
\citep{pogorelov_etal:11a,florinski_etal:04a}, or the Sun happens to be at rest relative to the 
interstellar medium \citep{czechowski_grygorczuk:17a} (see Appendix C).

Observations of energetic neutral atoms (ENAs) over energies 5.2---55~keV from the Ion and Neutral Camera (INCA) on the Cassini mission have shown rapid 2---3~year time variations \citep{dialynas_etal:17a}, which appear roughly correlated with the solar cycle. These observed 2---3 year time-variations by INCA are interpreted as requiring a line-of-sight that is limited by the size of the heliosheath. Since the observed variations of ENAs from all directions seem to be correlated in time, the shape of the heliosphere is argued to be spheroidal (i.e., round). Like the ``two-funnel'' heliosphere, this round heliosphere represents a significant departure from a comet-like shape. 

However, the interpretation of INCA data relies heavily on approximate equality between the upwind and downwind ENA flux intensity values, and the correlation between time variations of ENA fluxes from the upwind and downwind regions \citep{dialynas_etal:17a}, which we have found to be consistent with a comet-like shape.  Our simulations (Figures \ref{ftn}, \ref{fcmp}) show that the upwind and downwind ENA fluxes are, in fact, approximately equal to each other in the comet-like heliosphere, provided that the ENA energy is low enough. The time correlation between them can be explained within the comet-like paradigm of the heliosphere \citep{schwadron_bzowski:18a}. According to this explanation, the correlation is a consequence of time variations within the inner heliosheath driven by ram pressure changes in the SW and episodic cooling and heating of the inner-heliosheath plasma during the intervals of large-scale expansion and compression.

We conclude that the available observations of the directional distribution of the heliospheric 
ENA are qualitatively consistent with a comet-like structure of the heliosphere, of which the 
Huntsville model is an example. The approximate symmetry between the nose and the heliotail 
direction of ENA fluxes at energies up to several tens of keV is naturally explained by the 
decreasing magnitude of charge exchange cross section with energy. For increasing ENA energy, the 
tail to nose flux ratio is expected to increase. The future observations of very high energy ENA 
(up to $\sim$500 keV) by IMAP Ultra \citep{mccomas_etal:18a} may, therefore, provide a crucial test 
of the existence of the heliotail. The agreement between our model results and the actually 
observed ENA fluxes over a wide energy range (from a few keV to almost 100 keV) supports the 
scenario where the PUIs reflected, accelerated, and transmitted at the TS, as proposed by 
\citet{zank_etal:96b, zank_etal:10a, lee_etal:96a}, are indeed the source of the heliospheric 
ENAs.

\acknowledgements{A.C., M.B., and E.J.Z. acknowledge the collaboration within ISSI Team 368 \emph{The Physics of the Very Local Interstellar Medium and its interaction with the heliosphere}. A.C., M.B., J.M.S., and M.A.K. were supported by Polish NCN grants 2015-18-M-ST9-00036 and 2015-19-B-ST9-01328. G.P.Z., N.V.P., and J.H. acknowledge partial support through an IBEX subcontract to UAH. J.H. and N.V.P. were supported, in part, by NASA grants NNX15AN72G, NNX16AG83G, and 80NSSC18K1649NS.}

\appendix

\section{Parameter values for the simulation of the global heliosphere and the PUI density at 
the termination shock}
\noindent
In this section we demonstrate that parameter values adopted in the modeling are used 
consistently throughout the entire simulation process, starting from the unperturbed VLISM at the 
interstellar side and SW and solar EUV output at the solar side, and that they were chosen 
based on published up-to-date measurement values.

\subsection{Parameters of the Very Local Interstellar Medium obtained from heliospheric observations}
\noindent
The velocity of the Sun relative to the VLISM, i.e., the inflow velocity of interstellar matter 
on the heliosphere, and the VLISM temperature were adopted based on analysis of interstellar 
helium observations compiled by \citet{mccomas_etal:15b} $(T = 7500$ K, $v = 25.4$ km s$^{-1}$, 
ecliptic longitude $255.7\degr$, ecliptic latitude $5.1\degr$). The magnitudes of these 
quantities are based on analysis of direct-samplinig observations by IBEX-Lo from 2009--2014 
\citep{bzowski_etal:15a, schwadron_etal:15a, mobius_etal:15b}. These analyses do not rely on 
any particular model of the heliosphere and are based on the ballistics of neutral He atoms 
inside the heliosphere and a realistic model of ionization losses of interstellar He atoms 
inside the heliosphere. It is assumed that neutral He, H, and the plasma in the VLISM are in 
equilibrium. This assumption has been commonly made in the heliospheric physics. The 
temperature and flow velocity of the VLISM we used are robust against independent analyses of 
observations from Ulysses \citep{bzowski_etal:14a,wood_etal:15a} and the inflow direction 
obtained from pickup ion observations \citep{gloeckler_etal:04a, mobius_etal:15c}.

The density of interstellar neutral H at the upwind point of the termination shock was obtained 
based on two independent estimates: (1) the magnitude of slowdown of the SW expansion speed due 
to mass-loading by charge exchange with interstellar neutral H, observed in situ by Voyager 2 
\citep{richardson_etal:08a}, and (2) in situ measurement by Ulysses of the production rate of 
pickup ions at the boundary of the density cavity of interstellar H \citep{bzowski_etal:08a}. 
It was shown \citep{bzowski_etal:09a} that these estimates agree within mutual error bars. 
These authors suggested that the density of interstellar H at the nose of the termination shock 
is 0.09~cm$^{-3}$. With this, based on global MHD-kinetic modeling of the heliosphere done 
using the Moscow MC Model \citep{izmodenov_etal:03a}, the density of neutral H in the VLISM was 
estimated at 0.16~cm$^{-3}$, and the electron density $\sim0.06$~cm$^{-3}$.

The directions of inflow of the primary and secondary populations of ISN H were adopted in 
agreement with those for the primary and secondary populations of interstellar neutral He. The 
densities of these populations at the TS were adopted based on analysis of pickup ions observed 
by Ulysses \citep{bzowski_etal:08a}. The magnitudes of the parameters of the primary and 
secondary populations of ISN H we used are adopted after \citet{kowalska-leszczynska_etal:18b}.

The directions of inflow of the primary and secondary populations define the so-called neutral 
gas deflection plane (NDP). Heliospheric models suggest that the direction of inflow of the 
secondary population of neutral interstellar gas is located within the plane defined by the 
vectors of Sun's motion through the VLISM and the interstellar magnetic field (the B-V plane). 
The NDP found from observations of interstellar He (see Figure 7 in \cite{kubiak_etal:16a}) 
agrees very well with the B-V plane found from fitting the Ribbon size and location 
\citep{zirnstein_etal:16b}. In the present simulation of the heliosphere, we used the 
parameters obtained by these authors, listed in their Table 3 for the case of VLISM B field 3 
$\mu$G. For the B field, we used the direction and strength reported by these authors as 
resulting from ribbon fitting, i.e., $B = 2.93 \mu$G, $\lambda_B = 227.28\degr$, $\beta_B = 
34.62\degr$ at 1000 au ahead of the Sun. This vector is in agreement with an independent 
determination \citep{frisch_etal:15c} based on examination of the direction of polarization of 
starlight on interstellar dust grains.

\subsection{Solar wind and EUV conditions in the heliosphere and the sources of information on them}
\noindent
The parameters of the SW and solar EUV output used in the modeling of interstellar H 
and pickup ions were based on measurements (see Section 1). 

The model of SW speed and density during the solar cycle was adopted based on 
interplanetary scintillation observations outside the ecliptic plane \citep{tokumaru_etal:10a, 
tokumaru_etal:12b}, available from 1985, and within the ecliptic plane on in-situ data from the 
OMNI collection \citep{king_papitashvili:05}, compiled into a homogeneous model 
\citep{sokol_etal:13a,bzowski_etal:13a}. The SW density variations with heliolatitude 
were calculated based on a linear correlation between speed and densities at various latitudes, 
obtained from Ulysses in situ observations \citep{sokol_etal:13a} and the SW latitudinal invariant,
as described in \citet{leChat_etal:12a,sokol_etal:15d}. The photoionization rate was 
defined in Equations 3.23--3.25 from \citet{bzowski_etal:13a} based on EUV 
observations of the Sun and a system of solar proxies. The electron-impact ionization rate 
(important only within $\sim 1.5$ au from the Sun) was adopted from \citet{bzowski_etal:13a}.

The radiation pressure model was adopted from Equation 14 and Table 1 in 
\citet{kowalska-leszczynska_etal:18a}. The model was based on observations of the 
solar Lyman-$\alpha$ line profile during the solar cycle \citep{lemaire_etal:05} and the total 
irradiance in the Lyman-$\alpha$ line, available from the LASP Composite Line-Averaged Solar 
Lyman-$\alpha$ flux \citep{woods_etal:00}. In this model, the resonant radiation pressure acting on 
H atoms varies with time, heliolatitude, and atom radial velocity. 

These solar factors are calculated on a homogenous time grid, with averaging over Carrington 
period, and on a homogenous heliolatitude grid. The nWTPM model tracks individual atoms from a 
given location inside TS out to TS, and the solar factors are calculated along the trajectory, 
with their variation in time and with heliolatitude calculated by linear interpolation between the 
time- and heliolatitude nodes.

\section{Why is the nose/tail ENA flux ratio energy dependent: the toy model}
\noindent
A simple ``toy'' model of energetic ion distribution and production of ENAs in the heliosphere was 
proposed by \citet{czechowski_etal:18a}. Since the model offers a simple explanation for the energy 
dependence of the nose-tail ENA flux asymmetry in a "comet-like" heliosphere, a brief description
is included here. Two directions: the stagnation line (the 
"nose") and the center of the "tail" are considered. The ion distribution $J(z)$ along these 
directions is calculated taking into account the plasma convection towards the nose
(plasma speed $V(z)=V_0(1-z/L)$ and towards the tail (assuming plasma speed $V=$const) as well as the 
neutralization losses, with the loss rate $\beta_{cx}$. The ENA flux from the nose direction 
is then 
\begin{equation}
J_{ENA,nose}=\frac{\beta_{cx}L}{\beta_{cx}L+V_0}\frac{V_0}{v} J_{0,nose} 
\end{equation}
and from the tail 
\begin{equation}
J_{ENA,tail} = \frac{V}{v} J_{0,tail} 
\end{equation}
where $v$ is the speed 
of the ENA, $V$ the plasma speed in the tail, and $J_{0,nose}$, $J_{0,tail}$ are the energetic ion 
densities at TS in the respective directions. 

Assuming $L=25$ au, $V_{0,nose} = 100$ km s$^{-1}$, 
$V_{tail} = 26$ km s$^{-1}$, $J_{0,nose}/J_{0,tail} = 2$ (to account for the asymmetry of the 
termination shock), and taking the loss rate $\beta_{cx}=\sigma_{cx} n_H v$ where the ISN density
$n_H = 0.1$ cm$^{-3}$, and the charge-exchange cross section $\sigma_{cx}$ is given by the formula
of \citep{lindsay_stebbings:05a}, it follows that the ENA tail/nose flux at ratio 1 au is 1.6 
at the ENA energy 58 keV (the lowest HSTOF value) and becomes 1 at 46 keV (the midpoint of the 
highest INCA energy bin).

\section{MHD model of Parker solution for the heliosphere for the strong interstellar magnetic field}
\noindent
Available observations suggest that with the observed magnitudes of the plasma and magnetic field parameters in the VLISM the dominant component of the pressure balance is the ram pressure, not magnetic pressure \citep{schwadron_bzowski:18a} (although the magnetic pressure is high enough to cause some tilt and deformation of the heliotail). MHD models of the heliosphere are capable of reproducing solutions close to that presented in Figure 3 in the classical  paper \citep{parker:61a} provided that plasma flow velocity is almost nil (\citet{czechowski_grygorczuk:17a}, Figure 2) or directed parallel to the magnetic field direction \citep{pogorelov_etal:11a,florinski_etal:04a}. For inflow velocities not parallel to the magnetic field, as those obtained from analysis of neutral interstellar gas observations (20--25~km s$^{-1}$), and a very strong magnetic field of $20 \mu$G, the model of \citet{czechowski_grygorczuk:17a} predicts the plasma outflow mostly concentrated along two close-to-antiparallel funnels, but a heliotail is nevertheless present. The two-funnel structure gradually disappears for magnetic field strength decreasing towards the values available from observations, e.g., for $5 \mu$G in the unperturbed VLISM the funnels are absent (\citet{czechowski_grygorczuk:17a}, Figure 4).



\bibliographystyle{aasjournal}
\bibliography{iplbib}



\end{document}